\documentstyle[prl,aps]{revtex}

\input{tcilatex}

\begin{document}

\twocolumn[
\hsize\textwidth\columnwidth\hsize\csname@twocolumnfalse\endcsname
\title{\bf Ripplonic Polaron in the Quasi-One-Dimensional Electron System
over Liquid Helium}
\author{Sviatoslav S. Sokolov$^{1,2}$ and Nelson Studart$^1$}
\address{$^1$\it Departamento de F\'{\i }sica, Universidade Federal
de S\~{a}o Carlos,13565-905, S\~{a}o Carlos, S\~{a}o Paulo, Brazil}
\address{$^2$ B. I. Verkin Institute for Low Temperature Physics and 
Engineering, National Academy of Sciences of Ukraine, 310164, 
Kharkov, Ukraine}

\date{\today}
\maketitle

\begin{abstract}
We study the energetics and transport properties of polarons in the 
quasi-one dimensional electron system over the liquid helium surface.
 The localization lengths and the energy of the ground and excited states 
are calculated within the hydrodynamic model of the polaron. The polaron bound
 energy is found and the temperature for the polaron formation is estimated
 below 0.1 K. We examine the possibility of observation of the polarons by
 measuring both the spectroscopic transitions and the mobility.	 
\end{abstract}
\pacs{PACS numbers:  73.10.Di; 73.20.Dx; 73.90.+f}
]

An electron together with its self-induced polarization in a medium forms a
quasiparticle which has been named polaron. Besides its importance as a
standard theoretical model of a fermionic particle coupled to a boson scalar
field, the polaron has been observed in some physical systems. In
particular, the search of polaronic states, predicted theoretically long
time ago,\cite{theory-dimple,theory-froh} for surface electrons levitated
over liquid helium, has been the subject of large amount of experimental
work during the last decades.\cite{experiment,tress}

Theoretical approaches to investigate the properties of the surface polaron
over helium are based on the description of the dimple state (electron plus
the deformation of the helium surface due to the pressing field) within a
self-consistent hydrodynamic framework or in terms of the concept of a
Fr\"{o}hlich polaron (single electron coupled to ripplons). Despite the
great difference between the methods, both final results for the structure
of the ripplonic polaron show fair qualitative and quantitative agreement.%
\cite{review}

Quite recently, quasi-one-dimensional (Q1D) electron systems on the surface of liquid
helium  were realized.\cite{q1d-geo,q1d-eletr} In such a Q1D system,
the electron motion is restricted, in addition to the quantum well in the
direction normal to the liquid surface, by a lateral confinement, which is
provided by either geometric or electrostatic means, and can be modelled by
a parabolic potential. In such conditions, the formation of an {\it %
asymmetric }ripplonic polaron may become favorable.

In this work, we address the possibility of polaron formation in the Q1D
electron system within the hydrodynamic approach\cite{review,monarkha75,gs89}
which allows us to calculate analytically the main properties of the polaron
state. Furthermore, this model was employed by Tress et {\it al}. to
interpret the magnetoconductivity data which led them to report the
observation of polaronic effect in two-dimensional (2D) electron system over
helium films.\cite{tress} In the Q1D model, described in Refs. \cite
{km86,shs95}, the surface electrons are confined by a lateral potential well 
$U_{0}=m\omega _{0}^{2}y/2$ in the $y$ direction with characteristic frequency $%
\omega _{0}=\sqrt{eE_{\perp }/mR}$, where $e$ and $m$ are the electron
charge and mass, respectively, $E_{\perp }$ is the holding electric field
along the $z$ axis, and $R$ is the curvature radius of the liquid in the
channel.

The wave function $\psi (x,y)$ for the electron trapped to
the dimple satisfies the 2D Schr\"{o}dinger equation 
\begin{eqnarray}
-\frac{\hbar ^{2}}{2m}\left( \nabla ^{2}\psi \right) +\frac{m}{2}\omega
_{0}^{2}y^{2}\psi +eE_{\perp }\xi (x,y)\psi =\varepsilon \psi {,}\label{1}
\end{eqnarray}
where the deformation profile of the surface $\xi (x,y)$ depends, in
turn, on the electron pressure on the surface $eE_{\perp }|\psi (x,y)|^{2}$.
The total energy of the polaron, expressed as a sum of the electron energy
and the energy due to the surface deformation, is 
\begin{eqnarray}
W=\varepsilon +\frac{\sigma }{2}\int \int \left[ \left( {\bf \nabla }\xi
\right) ^{2}+k_{c}^{2}\xi ^{2}\right] dxdy{,}  \label{2}
\end{eqnarray}
where $k_{c}=\sqrt{\rho g/\sigma }$ is the capillary constant and $\sigma $
and $\rho $ are the surface tension coefficient and the density of liquid helium
respectively. The electron energy can be written as 
\begin{eqnarray}
\varepsilon  &=&\int \int \{\frac{\hbar ^{2}}{2m}\left[ \left( {\bf \nabla }%
\psi \right) ^{2}\right] +\frac{m}{2}\omega _{0}^{2}y^{2}\psi ^{2}  \nonumber
\\
&&+eE_{\perp }\xi (x,y)\psi ^{2}\}dxdy{,}  \label{3}
\end{eqnarray}
where we supposed that the characteristic length, which gives the decrease
of the wave function along the $y$ direction, is significantly smaller than
the curvature radius $R$. The minimization of the total energy $W$ with
respect to $\xi $ leads to the mechanical equilibrium equation 
\begin{eqnarray}
\sigma (\nabla ^{2}\xi -k_{c}^{2}\xi )=eE_{\perp }\psi ^{2}{.}  \label{4}
\end{eqnarray}
Equations (\ref{1}) and (\ref{4}) form a nonlinear system of equations which
must be solved self-consistently, as was done numerically in Ref.\cite{gs89}
for the 2D symmetric polaron. Since these calculations are cumbersome in the
asymmetric situation, we prefer to use a variational method and take trial
functions like the 2D harmonic-oscillator eigenfunctions with localization
lengths $l_{x}$ and $l_{x}$ as the variational parameters.\cite{monarkha75}
The result for the polaron ground-state energy reads as 
\begin{eqnarray}
W_{0} &\simeq &-\frac{(eE_{\perp })^{2}}{4\pi \sigma }\ln \frac{2\sqrt{2}}{%
\sqrt{\gamma }k_{c}(l_{x}+l_{y})}  \nonumber \\
&&+\frac{\hbar ^{2}}{4m}\left( \frac{1}{l_{x}^{2}}+\frac{1}{l_{y}^{2}}%
\right) +\frac{m}{4}\omega _{0}^{2}l_{y}^{2}{,}  \label{5}
\end{eqnarray}
where $L_{F}^{2}=(2\pi \sigma \hbar ^{2})/m(eE_{\perp })^{2}$ and $\gamma $
is the Euler-Mascheroni constant. The variational parameters are obtained
from the condition that the polaron energy is a minimum at these values,
which leads to the following system of equations: 
\begin{eqnarray}
\frac{1}{l_{x}^{3}}-\frac{1}{L_{F}^{2}(l_{x}+l_{y})}=0,  \label{6}
\end{eqnarray}
and 
\begin{eqnarray}
\frac{1}{l_{y}^{4}}-\frac{1}{L_{F}^{2}l_{y}(l_{x}+l_{y})}-\frac{1}{L_{0}^{4}}%
=0{,}  \label{6b}
\end{eqnarray}
where $L_{0}^{2}=\hbar /m\omega _{0}.$ If $L_{F}\ll L_{0}$, the solutions of
Eqs. (\ref{6}) and (\ref{6b}) are $l_{x}=l_{y}\simeq \sqrt{2}L_{F}$. For $%
L_{F}\gg L_{0}$, we obtain $l_{x}\simeq L_{F}$ and $l_{y}\simeq L_{0}.$ Note
that for realistic values of holding fields $L_{F}$ $\sim
10^{-1}$ $\mu $m and $L_{0}\sim 10^{-2}$ $\mu $m and the values of the
localization lengths are significantly smaller than the curvature radius $%
R=1-10$ $\mu $m. For the lowest excited states, we find 
\begin{eqnarray}
W_{10} &\simeq &-\frac{(eE_{\perp })^{2}}{4\pi \sigma }\left[ \ln \frac{2%
\sqrt{2}}{\sqrt{\gamma }k_{c}(d_{x}+d_{y})}-\frac{d_{x}(2d_{x}+3d_{y})}{%
4(d_{x}+d_{y})^{2}}\right]   \nonumber \\
&&+\frac{\hbar ^{2}}{4m}\left( \frac{3}{d_{x}^{2}}+\frac{1}{d_{y}^{2}}%
\right) +\frac{m}{4}\omega _{0}^{2}d_{y}^{2}{.}  \label{7}
\end{eqnarray}
The minimization of the energy with respect to the parameters $d_{x}$
and $d_{y}$ leads to

\begin{eqnarray}
\frac{3}{d_{x}^{3}}-\frac{1}{L_{F}^{2}(d_{x}+d_{y})}\left[ 1+\frac{%
d_{y}(d_{x}+3d_{y})}{4(d_{x}+d_{y})^{2}}\right] =0
\end{eqnarray}
and 
\begin{eqnarray}
\frac{1}{d_{y}^{4}}-\frac{1}{L_{F}^{2}d_{y}(d_{x}+d_{y})}\left[ 1-\frac{%
d_{x}(d_{x}+3d_{y})}{4(d_{x}+d_{y})^{2}}\right] -\frac{1}{L_{0}^{4}}=0{.}
\label{8}
\end{eqnarray}
In the limit of $L_{F}\ll L_{0}$, we can disregard the term $L_{0}^{-4}$ in
Eq. (\ref{8}) and the roots of the system are $d_{x}\simeq 2.12L_{F}$ and $%
d_{y}\simeq 1.73L_{F}$. In the opposite limit, $L_{0}\ll L_{F}$, we obtain $%
d_{x}\simeq \sqrt{3}L_{F}$ and $d_{y}\simeq L_{0}$.

We now investigate the transport properties of the Q1D polaron. When a
driving electric field $E_{\Vert }$ is applied along the plane, the surface
deformation moves together with the trapped electron inducing a field of
hydrodynamic velocities in the liquid, which is accompanied by energy
dissipation, and leads to a finite value of the polaron mobility.\cite
{theory-dimple,review,monarkha75,gs89} In order to evaluate the mobility, we
employ the energy balance equation $eE_{\Vert }v_{0}=d\rho _{E}/dt,$ where $%
v_{0}$ is the liquid velocity at infinity, and $\rho _{E}$ is the energy
density dissipated. The function $d\rho _{E}/dt$ is obtained in a
straightforward way by finding the normal velocity field induced by the
polaron from the solution of the Navier-Stokes equation.\cite
{theory-dimple,monarkha75} If the driving field is applied along the $x$
direction, the mobility is given by 
\begin{eqnarray}
\mu _{x}=\frac{e}{2\eta S}\left[ \sum_{\vec{k}}kk_{x}^{2}|\xi _{\vec{k}%
}|^{2}\right] ^{-1},  \label{9}
\end{eqnarray}
where $\eta $ is the helium viscosity, $S$ is the area of the liquid
surface, and $\xi _{\vec{k}}$ is the Fourier transform of the dimple
profile. The summation in Eq.(\ref{9}) can be performed analytically
resulting in a cumbersome function of complete elliptic integrals depending
on $l_{x}$ and $l_{y}.$ However, we can simplify significantly the
calculation by considering the limiting case $l_{x}\gg l_{y}$, which
corresponds to holding fields of interest $1-3\times (10^{6}$ V/m). In this
case, the polaron mobility can be written as 
\begin{eqnarray}
\mu _{x}=\frac{\pi ^{3/2}\sigma ^{2}l_{x}}{\sqrt{2}\eta eE_{\perp }^{2}}.
\label{10}
\end{eqnarray}
If the field is applied along $y$, the mobility is such that ($\mu _{x}/\mu
_{y})=\ln (l_{x}/l_{y})$. In the symmetric case, $l_{x}=l_{y}=L_{F}$, we
obtain the well-known result $\mu =\sqrt{8\pi }\sigma ^{2}L_{F}/\eta
eE_{\perp }^{2}.$

Given the recent experimental progress,\cite{tress,q1d-geo,q1d-eletr} we now
feel compelled to examine the signatures that should indicate the
experimental observation of the polaron in the Q1D electron system on the
surface of liquid helium. First, we calculate the polaron binding energy,
which we define as the ground-state polaron energy $W_{0}$ minus the energy
of the free electron in the Q1D channel. At low temperatures, $T\ll \hbar
\omega _{0}$, the electron energy in the lowest subband along the $y$
direction is $\hbar \omega _{0}/2.$ As we have seen previously, the values
of the variational localization lengths are $l_{x}\simeq L_{F}$ and $%
l_{y}\simeq L_{0}$ for $E_{\perp }=1-3\times (10^{6}$ V/m). In this case,
the binding energy is given by 
\begin{eqnarray}
E_{b}\simeq -\frac{(eE_{\perp })^{2}}{4\pi \sigma }\ln \frac{2\sqrt{2}}{%
\sqrt{\gamma }k_{c}L_{F}}+\frac{\hbar ^{2}}{4mL_{F}^{2}},  \label{11}
\end{eqnarray}
which is slightly smaller than the binding energy of the 2D symmetric
polaron. \cite{monarkha75,gs89} We estimate $E_{b}\simeq -0.03$ K and $%
E_{b}\simeq -0.3$ K for holding electric fields of $10^{5}$ V/m and $3\times
10^{5}$ V/m, respectively, in the case of a $^{4}$He substrate. One can
conclude that the energetic conditions for the formation of the polaron in
the Q1D electron system on liquid helium surface are near the same as in the
case of the 2D symmetric polaron on a flat helium surface and the Q1D
polaron should be observed at $T\lesssim |E_{b}|.$

Other possible way to observe the Q1D polaron is by measuring the
spectroscopic transitions between the ground and excited states of the
single electron in the dimple. By determining numerically the electron
energies in the ground and excited states we obtain that the frequency $%
\omega _{10}$ of the spectroscopic transition between the ground-state and
the first excited-state varies from $9.8$ $\times 10^{8}$ Hz, for $E_{\perp
}=10^{5}$ V/m, to $8.0\times 10^{9}$ Hz for $E_{\perp }=3\times 10^{5}$ V/m.
These values were obtained for the Q1D system over $^{4}$He. We remark that,
in this range of electric fields, the characteristic frequency $\omega _{0}$
varies from $5.9\times 10^{10}$ to $1.0\times 10^{11}$ Hz. As a consequence,
the frequency $\omega _{10}$ is one order of magnitude smaller than the
transition frequency between subbands of the $y$ - confinement. This
circumstance can be favorable in experimental efforts to observe the polaron
state by measuring the frequency of spectroscopic transitions.

Another possibility is measuring directly the mobility as in Ref.\cite{tress}%
. Our results point out that the mobility of the Q1D polaron is $40\%$
smaller than the mobility of 2D polaron for the same holding field. Our
value of the Q1D polaron mobility, for $T=0.1$ K and $E_{\perp }=3\times
10^{5}$ V/m, is three orders of magnitude smaller than the mobility of 2D
electrons over a flat helium surface and of electrons moving freely along a
channel, $\sim 10^{2}$ m$^{2}$/Vs,\cite{shs95}, which may allow to
distinguish them easily.

While the above estimates show that our proposed mechanism for polaron
formation in Q1D systems on helium should be compatible with accessible
experiments, other comments are needed to corroborate or refute our
predictions. Of particular interest is the possibility to use liquid $^{3}$%
He instead $^{4}$He as the substrate for the confined Q1D electron system.%
\cite{kono95} Recall that for $^{3}$He, whose surface tension coefficient is
more than twice smaller than that of $^{4}$He, the binding energy $|E_{b}|$
can be substantially larger. Also the spectroscopic transition frequencies
are twice larger for $^{3}$He in comparison with $^{4}$He. Furthermore, the
description of the polaron in terms of a hydrodynamic viscous model may be
doubtful for pure $^{4}$He for temperatures below $1$ K, whereas the
hydrodynamic approach of viscous phenomena is well satisfactory in the case
of $^{3}$He for temperatures down to $\sim 0.1$ K. Other attractive
substrate should be also the low-temperature mixture of $^{3}$He and $^{4}$%
He.\cite{hallock} On the other hand, unfortunately, one cannot devise now a
mechanism for using, in the Q1D apparatus, a solid substrate below the
liquid helium to provide a larger effective holding field as in the case of
the 2D polaron on a liquid helium film\cite{tress}, and as a consequence
greatly enlarging the absolute value of $E_{b}.$

In conclusion we predict the possibility for the formation of the asymmetric
polaron in the Q1D channel over the liquid helium surface. The localization
lengths of the asymmetric polaron, and the ground and excited state energies
are determined as functions of the holding electric field. The numerical
estimates of the frequency of spectroscopic transitions and the
temperature for the polaron formation indicate that the polaron in the Q1D
channel can be observed at accessible temperatures. One hope that the
asymmetric polaron can be observed at low temperatures, probably at $T<0.1$
K, and modern experimental methods for achieving low and ultralow
temperatures would offer the possibility to observe the polaron in the Q1D
channel on the liquid helium surface.

This work was partially sponsored by the Funda\c{c}\~{a}o de Amparo \`{a}
Pesquisa do Estado de S\~{a}o Paulo (FAPESP) and the Conselho Nacional de
Desenvolvimento Cientifico e Tecnol\'{o}gico (CNPq), Brazil.

\end{document}